\documentclass{article}
\usepackage[left=1in,top=1in,right=1in,bottom=1in]{geometry}
\usepackage{amsmath}
\usepackage{amsbsy}
\usepackage{hyperref}

\usepackage{amsthm}
\usepackage{epsfig}
\usepackage{color}
\usepackage{setspace}

\newcommand{\B}{B}

\newcommand{\N}{\mbox{{\small\textsc{N}}}}

\newcommand{\IG}{\mbox{{\small\textsc{IG}}}}

\newcommand{\E}{\mbox{E}}

\newcommand{\I}{I}

\newenvironment{example}[1][Example]{\begin{trivlist}
\item[\hskip \labelsep {\bfseries #1}]}{\end{trivlist}}

\usepackage{bm}
\usepackage{natbib}

\title{Predictor-dependent shrinkage for linear regression via partial factor modeling} 

\author{P. Richard Hahn\\
Department of Statistical Science, Duke University\\
\\
Sayan Mukherjee\\ 
Departments of Statistical Science, \\
Computer Science, Mathematics, and \\
Institute for Genome Sciences 
\& Policy, Duke University\\
\\
Carlos Carvalho\\
McCombs School of Business, The University of Texas, Austin}
\date{}
\begin{document}
\maketitle 

\abstract
In prediction problems with more predictors than observations, it can sometimes be helpful to use a joint probability model, $\pi(Y,X)$, rather than a purely conditional model, $\pi(Y \mid X)$, where $Y$ is a scalar response variable and $X$ is a vector of predictors.  This approach is motivated by the fact that in many situations the marginal predictor distribution $\pi(X)$ can provide useful information about the parameter values governing the conditional regression.  However, under very mild misspecification, this marginal distribution can also lead conditional inferences astray.  Here, we explore these ideas in the context of linear factor models, to understand how they play out in a familiar setting.  The resulting Bayesian model performs well across a wide range of covariance structures, on real and simulated data.

\phantomsection
\doublespacing

\section{Introduction}
Consider regressing a scalar response $Y$ on a vector of predictors $X$, when the number of independent replications, $n$, is much smaller than the number of predictors, $p$. Assume that the goal is to provide reliable predictions along with associated confidence statements.  This paper focuses on the following question:  assuming that we know the form of the conditional distribution $\pi(Y \mid X, \beta)$, how should the marginal distribution of the predictors $\pi(X)$ inform our estimates of $\beta$?

Within a Bayesian framework one may pass information through a joint sampling model \citep{unlabeled}.  In the $n \ll p$ setting, a parsimonious assumption is that the covariation among the elements of $X$ and between $X$ and $Y$ can be captured by a lower dimensional set of latent variables, which we denote by $f$.  Generically this may be expressed as 
\begin{equation}
\begin{split}
\pi(Y, X \mid f, \beta) &=\pi(X \mid f)\pi(Y \mid f),
\end{split}
\end{equation}
where $k \equiv \mbox{dim}(f) \ll p$.  This structure describes conditional independence of $Y$ and $X$, given $f$.

While natural, this approach presents an often overlooked modeling challenge.  Because the sampling distribution for $X$ is of much higher dimension than the regression model, posterior inference on the latent factors $f$ is liable to be overwhelmingly determined by this marginal distribution, essentially ignoring $Y$.  When $k$ is chosen inadequately small, it may be mistakenly inferred that the response is entirely uncorrelated with the predictors.  The joint likelihood is dominated by $X$, even if our practical goal is to use $X$ to predict $Y$.  An analogous problem in principal component regression is well known;  the {\it least eigenvalue scenario} is when the response is associated strongly only with the least important principal component \citep{hotelling, cox, jolliffe82}. 

There are two common tactics for dealing with this problem.  The first is simply to use a conditional model.  This approach has the virtue of limiting the number of free parameters one must interpret and compute with.  It has the drawback that information about $X$ must be incorporated into the regression with no accompanying reliability assessment.  For example, in singular value regression, one takes the top $k \ll p$ left singular vectors of the design matrix as the predictors.  Such procedures do not propagate uncertainty about this choice of $k$ into predictions and confidence regions.  

The second approach is to place a prior on $k$, including it in a full Bayesian model, thus allowing inference on $k$.  Though this approach inherently propagates uncertainty about $k$, specifying a prior over $k$ that respects the goal of prediction within the framework of the joint distribution is nontrivial (see example 2, section 2.2).  

To fix ideas, this paper studies the above issues in a Normal linear regression setting, where
\begin{eqnarray}
(Y_i \mid X_i, \beta, \sigma) &\sim&\N(X^t_i\beta,\sigma^2).
\end{eqnarray}
As our marginal predictor model we study a Bayesian factor model \cite{West03},  
\begin{equation}\label{factormodel}
\begin{split}
X_i &= \B f_i+\nu_i, \hspace{0.1in} \nu\sim\N(0,\Psi) \\
f_i&\sim\N(0,\I_k).
\end{split}
\end{equation}
Without loss of generality we assume throughout that our response and predictor variables are centered at zero.

In the next section, we demonstrate the challenges of prior specification in this setting, in terms of obtaining a satisfactory conditional regression.  Rather than tackling this prior specification head on, our solution is to construct a hierarchical model which is centred at the Bayesian factor regression model. Permitting deviations from this model safeguards inference against sensitivity to the choice of the number of factors included in the model, sidestepping the intrinsic sensitivity to prior specification.

Section four compares our method to common alternatives, such as ridge regression, partial least squares, principal component regression \citep{elements} and least angle regression \citep{lars} on real and simulated data.  Principal components, partial least squares and least angle regression all explicitly incorporate features of the observed predictor space when making predictions, while ridge regression does not, a distinction which is further discussed in Section 3.2.  Section five considers extensions to variable selection and subspace estimation.  

\section{The effect of $k$ on factor model regression}
\subsection{Bayesian linear factor models}
We briefly provide details of a typical Bayesian linear factor model.  Any multivariate Normal distribution may be written in {\it factor form} as in (\ref{factormodel}).  The matrix $\B$ is a $p \times k$ real-valued matrix and $\Psi$ is diagonal.  The matrix $B$ is referred to as a loadings matrix, the elements of $\Psi$ are referred to as idiosyncratic variances, and the $f_i$ are called factor scores.  Conditional on $\B$ and $f_i$, the elements of each observation are independent. 
Integrating over $f$, we see 
\begin{equation}\label{decomp}
\mbox{cov}(X) \equiv \Sigma_X = \B\B^t +\Psi.
\end{equation}
When $k=p$ this form is unrestricted in that any positive definite matrix can be written as (\ref{decomp}).  We say that a positive definite matrix admits a $k$-factor form if it can be written in factor form $\B \B^t + \Psi$ where $\mbox{rank}(\B) \leq k$.  Note that $BB^t + \Psi$ has full rank whenever the idiosyncratic variances are strictly positive, while $B$, which encodes the covariance structure, may have much lower rank.

If we further assume that the $p$ predictors influence the response $Y$ only through the $k$-dimensional latent variable $f$, we arrive at the Bayesian factor regression model of \cite{West03}:
 \begin{equation}\label{freg}
 \begin{split}
Y_i &= \theta f_i + \epsilon_i, \hspace{0.1in} \epsilon\sim\N(0,\sigma^2)\\
\Sigma = \mbox{cov}\begin{pmatrix}X \\Y \end{pmatrix}&=\begin{bmatrix} \B\B^t+\Psi &\hspace{0.5cm} V^t\\
V& \;\;\;\omega \end{bmatrix}, \\
V&=\theta B^t,\\
\omega&=\sigma^2 + \theta \theta^t.
\end{split}
 \end{equation}     
As the norm of $\Psi$ goes to zero, this model recovers singular value regression.  Here $\theta$ is a $1 \times k$ row vector; effectively it is an additional row of the loadings matrix ($\theta \equiv b_{p+1}$ and $Y_i = X_{p+1,i}$).  

Factor models have been a topic of research for over a century, with increased recent interest spurred by the ready availability of computational implementations.  A seminal reference is \cite{spearman1904};  \cite{press:1982} and \cite{bartholomew} are key modern references.  Bayesian factor models for continuous data have been developed by many authors, including \cite{GewekeZhou96} and \cite{AguilarWest00}.  A thorough bibliography can be found in \citep{lopesbiblio}.   Notable applications include finance \citep{AguilarWest00, famafrench1,famafrench2,LV,Bai, chamberlain,chamberlainroth, lopescarvalho} and gene expression studies \citep{merl, lucas, carlosJASA}.  The area continues to see new methodological developments focusing on a variety of issues:  prior specification \citep{ghoshdunson}, model selection \citep{Lopes,anirban} and identification \cite{lopesnew}. In this work we highlight the use of factor models for prediction.  

\subsection{The effects of misspecifying $k$}
If $k$ is chosen too small, model inferences can be unreliable as a trivial consequence of misspecification.  Less appreciated, however, is that minute misspecifications in terms of overall model fit can drastically impair the suitability of the regression induced by the joint model.  The following two examples demonstrate that the evidence provided by the data may be indifferent between two factor models which differ only by the presence of one factor, even though the larger model is strongly preferred by some prediction criterion.  In the first example this can be observed analytically; the second example demonstrates this effect via simulation.

\begin{example}
Consider returns on petroleum in the United States and in Europe and assume we are interested in estimating the spread for trading purposes. Let $X = (X_1 , X_2 )$, where $X_1$ and $X_2$ is the price in the U.S. and in Europe, 
respectively, so that we want to predict $X_1- X_2$ . If we consider the correlation matrix, the first principal component will 
be given by $X_1 + X_2$ with variance $\frac{1+r}{\sqrt{2}}$ while the second component is $X_1 - X_2,$ with variance $\frac{1-r}{\sqrt{2}}$ where $r$ is the correlation between the two prices.  For $r$ near one, a regression based on only the first principal 
component will discard all the relevant information, because the second principal component is the one of interest \citep{forzaniunpub}.
\end{example}
We see that the bias incurred by throwing away the second principal component is much bigger than the reduction in variance incurred by its elimination.  

In the bivariate case, this discrepancy may seem inconsequential.  But with even a moderate number of predictors, deciding whether or not to add an additional factor can be difficult, as the next example illustrates.

\begin{example}
Consider the 10-dimensional two-factor Gaussian model with loadings matrix 
\begin{equation}
\begin{split}
\begin{align*}
\B^t &= \begin{bmatrix} 
     0 & \hspace{0.1in}   -4 &\hspace{0.1in}    \phantom{-}0 &\hspace{0.1in} -8 &\hspace{0.1in}-4 &\hspace{0.1in}   -6 &\hspace{0.1in}   \phantom{-}1 & \hspace{0.1in}-1 &  \hspace{0.1in}  \phantom{-}4 &  \hspace{0.1in}  \phantom{-}0\\
     1  &\hspace{0.1in} \phantom{-}0   &\hspace{0.1in}  \phantom{-}0   &\hspace{0.1in}-1   &\hspace{0.1in} \phantom{-}0    &\hspace{0.1in}\phantom{-}1   & \hspace{0.1in} \phantom{-}0   &\hspace{0.1in}  \phantom{-}1    &\hspace{0.1in}\phantom{-} 0   &\hspace{0.1in} \phantom{-}1
 \end{bmatrix}
   \end{align*}
 \end{split}
 \end{equation}
 and idiosyncratic variances $\psi_{jj}= 0.2$ for all $j \in \lbrace 1, \dots, p\rbrace$.   Now consider the one-factor model that is closest in KL-divergence to this model, with loadings matrix 
 \begin{equation}
\begin{split}
\begin{align*}
A^t &= \begin{bmatrix} 
    0.0004 &\hspace{0.1in}    -3.9967 &\hspace{0.1in}   0&\hspace{0.1in}    -7.9713   &\hspace{0.1in} -3.9967   &\hspace{0.1in} -5.9778 &\hspace{0.1in}    0.9990&\hspace{0.1in}    -0.9960   &\hspace{0.1in}  3.9967   &\hspace{0.1in} -0.0004
 \end{bmatrix}
   \end{align*}
 \end{split}
 \end{equation}
  and idiosyncratic variances given by the vector 
  \begin{equation*}
  D  = \begin{bmatrix} 
    1.2000 &\hspace{0.1in}  0.1871  &\hspace{0.1in}   0.2000 &\hspace{0.1in}     1.5032 &\hspace{0.1in}    0.1871  &\hspace{0.1in}   1.3762  &\hspace{0.1in}   0.1996   &\hspace{0.1in}  1.2054 &\hspace{0.1in}    0.1872    &\hspace{0.1in}  1.2000
 \end{bmatrix}.
  \end{equation*}
  Observe that the one-factor loadings matrix $A$ is very nearly equal to the first factor of $\B$, but that the idiosyncratic variances are notably different.  In particular, consider the problem of using the one-factor approximation to predict future observations of the $10^{\mbox{th}}$ dimension of $X$, which does not load on the first factor (similar to the first example).  The true idiosyncratic variance is $\psi_{10} = 0.2$, but the approximate model has $D_{10} = 1.2$, suggesting that prediction on this dimension will be inaccurate.  However, as measured by the joint likelihood, the one factor model is an excellent approximation.  These mismatched conclusions are reflected in the following graph, which plots the difference in mean-squared prediction error between the two models against the difference in log-likelihood; each point represents a realization of 10 observations.  Above zero on the vertical axis favors the true model, while below zero favors the one-factor approximation.  The horizontal axis represents approximation loss due to the missing factor.  The average likelihood ratio is approximately one, while prediction performance is always worse with the smaller model.
  
  \begin{figure}[h!]
\begin{center}
\includegraphics[width=3.5in]{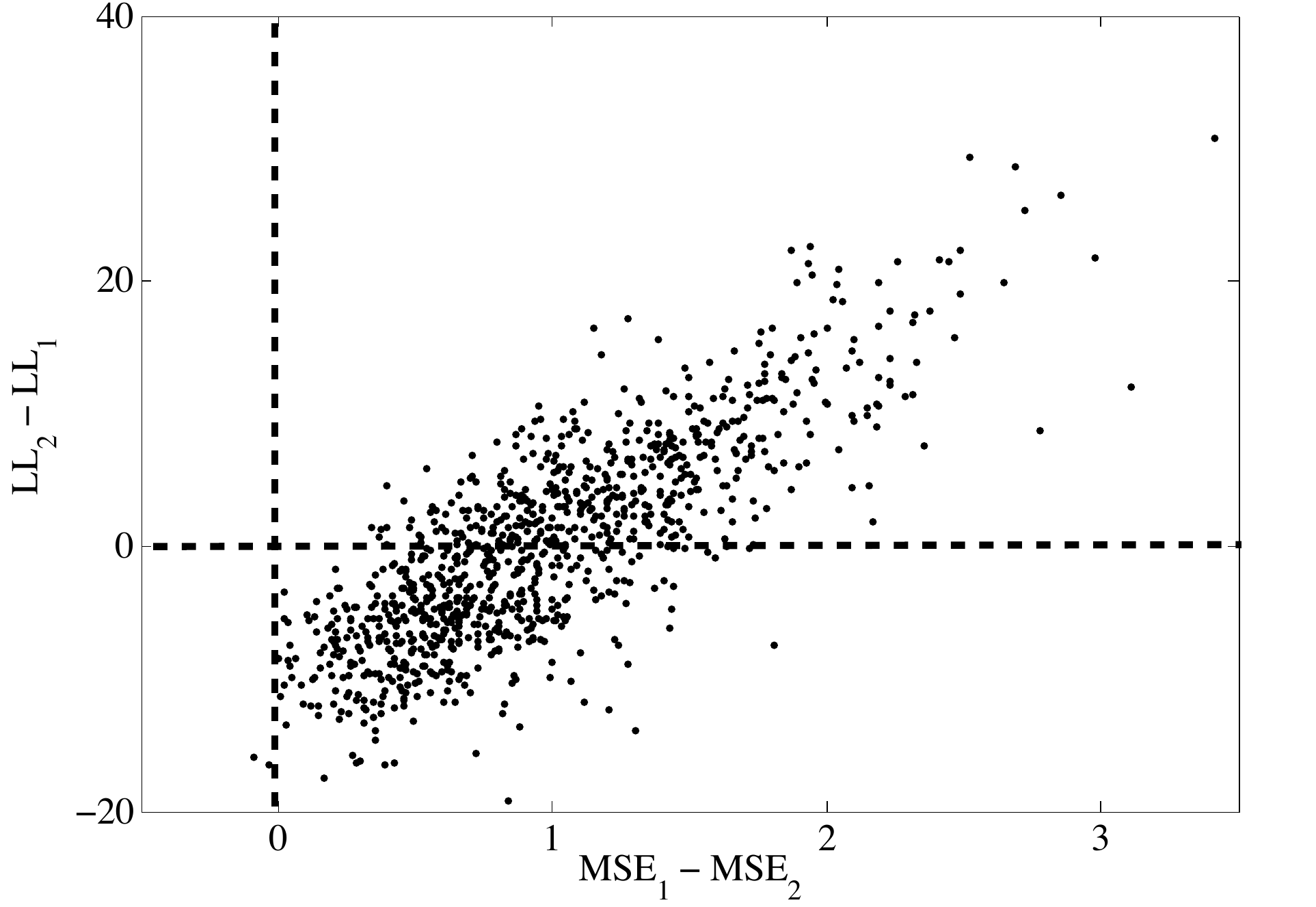}
\end{center}
\caption{Points denote realizations from the true two-factor model.  For points above the dashed horizontal line the likelihood ratio favors the true model.  The distance to the right of the dashed vertical line measures how much worse than the true model the one-factor approximation did in terms of predicting $X_{10}$. Model selection based on the full likelihood favors the larger model half the time, while model selection based on predictive fit favors the larger model nearly always.}
 \end{figure}
\end{example}

 More importantly, this discrepancy does not fade as we collect more data.  With only 10 observations, the likelihood ratio favors the true model only 47\% of the time;  with 100 observations this number creeps up to 51\% and at 1000 observations it stays at 51\%.  By the likelihood criterion the two models are virtually identical.  However, in terms of predicting $X_{10}$, the one-factor approximation is literally useless:  the conditional and marginal variances are virtually identical.  

Thus we see that relying on a prior distribution to correctly chose between a one- versus two-factor model is a difficult task: the prior would have to be strong enough to overwhelm more than a thousand observations worth of evidence which favors the wrong model about half the time.  

It may be instructive for some readers to understand this phenomenon from a matrix decomposition point of view, by defining $m$ to be the optimal value of the {\it Frisch problem}  \citep{frisch}:
  \begin{equation}
 \begin{split}
\mbox{minimize   } \mbox{rank}(\Sigma - \Psi)\\
\mbox{subject to  }\Sigma - \Psi &\succeq 0,
\end{split}
 \end{equation}   
 with $\Sigma$ a fixed covariance matrix and the optimization performed over $\Psi$, the diagonal idiosyncratic variance matrix;  $M \succeq 0$ denotes $M$ positive semi-definite.  This rank minimization problem is known to be NP-hard \citep{nphard}; this means, intuitively, that the minimum rank is very sensitive to small changes in $\Psi$.  This hardness implies, conversely, that for unknown $B$ of fixed rank $k$ and unknown $\Psi$, there exist matrices $\Sigma$ which we may approximate arbitrarily closely as $BB^t + \Psi$, although $m > k$. 
 
 By contrast, a cross-validation approach would uncover the predictive superiority of the two-factor model directly.   While a joint distribution allows one to borrow information from the marginal predictor distribution, which may be useful for prediction, using an unmodified high dimensional joint distribution subjugates the prediction task to the potentially more difficult task of high dimensional model selection.  These difficulties persist even with the use of sophisticated nonparametric model selection priors for factor models \citep{anirban}, because the trouble lies not with any particular prior, but rather with the assumption that the latent factors $f$ explain {\it all} of the variability in {\it both} $X$ and $Y$. 
 
In the next section we surmount the difficulty directly, by relaxing the assumption that the latent factors capturing the predictor covariance are also sufficient for predicting the response.
n letting $k$ be inferred, if the priors over all of the parameters of the model are not carefully chosen.  

\section{Partial factor regression}
\subsection{Specification}
Our new model --  referred to here as the {\it partial factor} model -- circumvents the prior specification difficulties described in the previous section by positing a lower-dimensional covariance structure for the predictors, but permitting the relationship between the predictors and the response to be linear in up to $p$ dimensions.  This is achieved by using the following covariance structure for the joint Normal distribution:
\begin{equation}\label{pfr}
\begin{split}
\begin{pmatrix}X\\ Y \end{pmatrix} &\sim\N(0, \Sigma)\\
\Sigma&=\begin{bmatrix} BB^t+\Psi & \hspace{0.5cm} V^t\\
V& \omega \end{bmatrix}.
\end{split}
\end{equation}
The difference between (\ref{freg}) and (\ref{pfr}) is that in (\ref{pfr}) $V$ is not required to equal $\theta B^t$.  The matrix $B$ is still a $p \times k$ matrix with $k\leq n\ll p$ so that the $p \times p$ predictor covariance matrix is constrained to the $BB^t+\Psi$ form, but, the full covariance matrix $\Sigma$ is not simultaneously restricted.  This way, the response can depend on directions in predictor space which are not dominant directions of variability,  but inference and prediction still benefit from this structural regularization of $\Sigma_X$. 

Just as crucially, the prior on $V$ may be conditioned on $\Sigma_X$.  Specifically, we may suggest, via the prior, that higher variance directions in predictor space are more apt to be predictive of the response.  But, unlike principal component regression or factor models, the prior furnishes this bias as a hint rather than a rigid assumption; hints are important in $p>n$ settings.   

The hierarchical specification arises from the jointly Normal distribution between $X$, $Y$, and the $k$ latent factors, which have covariance:
\begin{align}\label{fullcov}
\mbox{cov}\begin{pmatrix}X \\ f\\Y \end{pmatrix}&=\begin{bmatrix} BB^t+\Psi \;\;&\hspace{0.5cm} B^t \hspace{0.5cm} &V^t\\
B& \I_k&\theta^t \\
V&\theta&\omega
 \end{bmatrix}.
\end{align}
From this covariance, the conditional moments of the response can be expressed as:
\begin{eqnarray}\label{expect}
E(Y \mid f, X)&=&\theta f+\{(V-\theta B^t)\Psi^{-\frac{1}{2}}\}\{\Psi^{-\frac{1}{2}}(x- B f)\}\\
\mbox{var}(Y \mid f, X)&=&\omega - [V \hspace{.1in}\theta]\Sigma_{X,f}^{-1}[V \hspace{.1in}\theta]^t \hspace{.1in}\equiv \sigma^2.
\end{eqnarray}
A natural prior for $V$, conditional on $\theta$, $\B$ and $\Psi$ might be $$V \sim \N(\theta B^t, w^{-1}\Psi),$$ implying that a priori the error piece plays no role in the regression.  A reasonable choice of independent Normal prior on $\theta$ would be $$\theta \sim \N(0, \I_k),$$ because the scale of the factors are set to have unit variance.  All together, the model may be expressed as
\begin{equation}\label{pfmod}
\begin{split}
X\mid B, f,\Psi &\sim\N(\B f, \Psi)\\
Y \mid X, B, \theta, V, f, \Psi, \sigma^2 &\sim\N(\theta f+\{(V-\theta B^t)\Psi^{-\frac{1}{2}}\}\{\Psi^{-\frac{1}{2}}(x- B f)\},\sigma^2)\\
V\mid \theta, B, \Psi&\sim\N(\theta B^t, w^{-1}\Psi ),\\
f&\sim\N(0,\I_k)\\
\theta\sim&\N(0,q^{-1}\I_k)\\
b_{jg}\mid\psi_j,\xi_g&\sim\N(0, \xi_g^{-1}\psi_j),  \; \; g = 1, \dots, k,\;\; j=1,\dots,p.
\end{split}
\end{equation}

The conditional regression parameters now borrow information from the marginal distribution via the prior -- we have centered the regression at the pure factor model.  However, the data may steer us away from this assumption. By decoupling the predictor distribution from the conditional distribution, prior specification on the potentially ultra-high dimensional predictor space does not affect our lower dimensional regression in counterproductive ways.  At the same time, the hierarchical prior on the regression parameters facilitates the borrowing of information that is necessary in the $p \gg n$ setting.
\subsection{A conditional distribution view}

Note that the prior on $V$, marginalizing over $\theta$,  is
\begin{equation}
V \mid B, \Psi \sim \N(0, BB^t+\Psi) = \N(0,\Sigma_x).
\end{equation}
 Because $\beta = V\Sigma_X^{-1}$,
\begin{equation}
\begin{split}
\mbox{cov}(\beta)&=\Sigma_X^{-1}\Sigma_X\Sigma_X^{-1}\\
&=\Sigma_X^{-1}.
\end{split}
\end{equation}  
In other words, the partial factor model is a special case of the following hierarchical model:
\begin{equation}\label{hiermod}
\begin{split}
X_i \mid \Sigma_X &\sim \N(0, \Sigma_X)\\
Y_i \mid X_i, \beta, \sigma^2 &\sim \N(X_i^t\beta, \sigma^2)\\
\beta \mid \Sigma_X, \tau, \sigma^2 &\sim \N(0, \tau^{-1} \sigma^2 \Sigma_X^{-1})
\end{split}
\end{equation}
where $\Sigma_X$ is restricted to have $k$-factor form.  Note that conditional on $\Sigma_X$ this is simply a conjugate Normal-Inverse-Gamma prior on the regression parameters:
\begin{equation}
\begin{split}
\beta \mid \sigma^2, S_0 &\sim\N(0,\sigma^2S_0)\\
\sigma^2 &\sim \IG(a,b),
\end{split}
\end{equation}
with $S_0 = \tau^{-1}\Sigma_X^{-1}$.  This observation permits easy comparison to two other common linear regression priors.
Taking the prior covariance matrix to be $S_0 = \tau^{-1} \I$  gives the well-known ridge estimator: 
\begin{eqnarray}
\tilde{\beta}\hspace{.1in} =\hspace{.1in} \E(\beta \mid Y, X)&=&(XX^t+ \tau\I_p)^{-1}XX^t\hat{\beta},
\end{eqnarray}
where $\hat{\beta}$ is the (generalized) least-squares estimator
$$\hat{\beta} = (XX^t)^{\dagger}XY$$ (where $M^{\dagger}$ denotes the Moore-Penrose pseudo-inverse \citep{golub1996} of $M$).  
Similarly, Zellner's $g$-prior \citep{zellner86, merlise} takes $S_0=g^{-1}(XX^t)^{\dagger}$ yielding the estimator
\begin{eqnarray}
\bar{\beta}\hspace{.1in} =\hspace{.1in} \E(\beta \mid Y, X) &=&(1+g)^{-1}\hat{\beta}.
\end{eqnarray}

To appreciate the benefit of using (\ref{hiermod}), consider the usual rationale behind the ridge regression prior versus that of the $g$-prior .   It is straightforward to show that the ridge estimator downweights the contribution of the directions in (observed) predictor space with lower sample variance, from which one may argue that \citep{elements}:  \begin{quote}
ridge regression protects against the potentially high variance of gradients estimated in the short directions.  The implicit assumption is that the response will tend to vary most in the directions of high variance in the inputs.
\end{quote}
The $g$-prior, by contrast, shrinks $\beta$ more in directions of high sample variance in the predictor space a priori, which has the {\it net effect} of shrinking the orthogonal directions of the design space equally regardless of whether the directions are long or short.   This reflects the {\it substantive} belief that higher variance directions in predictor space need not influence the response variable more than the directions of lower variance.

However, this story conflates the observed design space with the pattern of stochastic covariation characterizing the random predictor variable.  It would be more desirable to realize the benefit of regularizing estimates in directions of low sample variance, while not over-regularizing regions of predictor space with weak stochastic covariance structure.  Teasing apart these two aspects of the problem can be done by conditioning on $X$ and $\Sigma_{_X} \equiv \mbox{cov}(X)$ separately, exactly as (\ref{hiermod}) does.  

We may observe this teasing-apart effect directly from the form of the estimator under (\ref{hiermod}).  Assuming for simplicity that $\lambda  = \sigma^2 =1$, and let $\hat{\Sigma}_X \equiv n^{-1}XX^t$ and $\hat{V}\equiv n^{-1}XY$.  Then
\begin{eqnarray}\label{stochastic}
\E(\beta \mid Y, X, \Sigma_X) &=&(\I_p + n \Sigma_x^{-1}\hat{\Sigma}_X)^{-1}(\Sigma_X^{-1}V_0 + n \Sigma_X^{-1}\hat{V}),\\
\beta_0&=&\Sigma_X^{-1}V_0,
\end{eqnarray}  
where $V_0$ is chosen a priori and determines the prior mean of the regression coefficients.  Because $\Sigma_X$ and $\hat{\Sigma}_X$ are never identical, we still get shrinkage in different directions, thus combatting the ``high variance of gradients estimated in short directions" while not having to assume that any direction in predictor space is more or less important a priori.

In this light, we see that ridge regression is motivated by a mathematical fact about regularization, while the $g$-prior is motivated by a substantive belief regarding the influence of the predictor variables on the response (namely, symmetry).  The partial factor model can be understood as using $X$ to learn about $\Sigma_X$ and then using this information when trying to learn $\beta$.  

Moreover, Zellner's $g$-prior may be interpreted as a crude approximation to this idea -- rather than as a misguided regularization tool that shrinks the impact of reliably measured covariates more than unreliable ones.  The crucial distinction is whether or not the predictors are taken to be fixed or stochastic.  For example,  \cite{maruyama}  advocate ``more shrinkage on higher variance estimates" and construct a prior on $\beta$ which involves $X$, much like the $g$-prior, but which amplifies the effect of ridge regression in that it results in more shrinkage in observed directions of low sample variance. However, in the case of stochastic predictors, one must distinguish between $XX^t/n$ and $\Sigma_X$, as we have seen in (\ref{stochastic}).  The partial factor model, which centres the conditional regression at a low-dimensional factor model, actually recovers the $g$-prior-esque (\ref{hiermod}).  However, the $k$-factor structure imposed on $\Sigma_X$ by the partial factor model provides a much improved estimator of $\Sigma_X$ over the naive sample covariance estimate that appears in the $g$-prior.
	
It is further instructive to consider the case where $\Sigma_X$ is given.  Here, the difference between (\ref{hiermod}) and ridge regression amounts to placing an independent prior on the regression coefficients associated with the de-correlated predictors as opposed to those corresponding to the original -- possibly correlated -- predictors.   To see the equivalence, let $\tilde{X} = (L^{t})^{-1}X$, where $L^{t}L=\Sigma_X$ is the Cholesky decomposition of the covariance matrix so that
\begin{equation}
\begin{split}
\begin{pmatrix}\tilde{X} \\ Y \end{pmatrix} &\sim\N(0, \tilde{\Sigma}),\\
\tilde{\Sigma}&=\begin{bmatrix} \I_p & \hspace{0.5cm} \alpha^t\\
\alpha& \omega \end{bmatrix}.
\end{split}
\end{equation}
Then an independent prior on this regression $(\alpha \mid \sigma^2, \tau) \sim \N (0,\sigma^2\tau^{-1} \I_p)$ implies $$(\beta \mid \Sigma_X, \tau, \sigma^2) \sim \N(0,\sigma^2\tau^{-1}\Sigma_X^{-1})$$ as in (\ref{hiermod}) above.  

This simple observation raises interesting questions about the role of ``sparsity" in linear regression models with stochastic predictors.  Indeed, believing it plausible that some of the regression coefficients are identically zero is incompatible with the assumption that the same is true of the coefficients in the de-correlated predictor space (for arbitrary covariances).  

\subsection{Efficient approximation}
Sampling from the posterior distribution of the partial factor model may be achieved via standard Markov chain Monte Carlo methods.  In particular, a Gibbs sampler for the ordinary factor model provides an excellent proposal distribution for a Metropolis-Hastings update for many of the parameters.  This approach provides measures of posterior uncertainty over all parameters, up to Monte Carlo error.  This approach is slow, however, owing to the need to compute the determinant of a $p$-dimensional matrix in computing the acceptance ratio.  For the purpose of prediction, the following approximation, which we call {\it partial factor regression}, proves useful.
 
Partial factor regression applies ridge regression to an augmented design matrix with elements 
\begin{equation}
\begin{split}
Z_i &= [f_i \; \; \; r_i]\\
r_i &= (X_i - Bf_i)\Psi^{-\frac{1}{2}}
\end{split}
\end{equation}
mimicking the expression in (\ref{expect}).  {\it Two} regularization parameters, $\tau_f$ and $\tau_r$, are then selected by cross-validation, corresponding to the respective regression coefficients on the latent factors and the residuals; these are analgous to $q$ and $w$ in (\ref{pfmod}) .  Point estimates are obtained for $Z_i$ as the posterior mean of (\ref{factormodel}) using a Gibbs sampling implementation.  Partial factor ridge regression may be written as 
\begin{equation}
\begin{split}
Y_i \mid \hat{Z}_i, \gamma, \sigma^2 &\sim \N(\hat{Z}_i^t\gamma, \sigma^2),\\
\gamma \mid \tau_f, \tau_r &\sim \N(0, \sigma^2 S_0),\\
S_0 &= \begin{bmatrix}\tau_f^{-1} \I_k  \hspace{0.5cm} & 0\\
0& \tau_r^{-1} \I_p  \end{bmatrix},\\
\hat{Z}_i & = \E(Z_i),
\end{split}
\end{equation}  
where the expectation in the last line is taken over the posterior $\pi(B, \Psi, f_i \mid X_{1:n})$ derived from model (\ref{factormodel}).

This approach ignores the impact of $Y$ on learning these parameters under the partial factor model; however, this contribution should be minor by the arguments of Section two, turning a model flaw in the factor modeling context into a computational shortcut in the partial factor setting. This step of the procedure may be done ahead of time and using as much marginal predictor data as is available, to better estimate $\Sigma_X$.  Aside from this preprocessing, the model fitting is exactly ridge regression using the augmented design matrix.

Moreover, this expression of the partial factor idea makes transparent where gains may be achieved over other methods -- by decomposing the regularization component into two separate pieces, one concerned with the marginal stochastic structure of the predictors and the other dealing directly with the conditional regression model.  

Viewed from this perspective, the partial factor model is an instantiation of the manifold regularization approach of \cite{niyogiMani}, but motivated by an underlying generative model;  $\tau_f$ is the ``intrinsic" penalty parameter and $\tau_r$ is an additional ``ambient" penalty parameter.  The key insight underlying the partial factor model is precisely that these two components may be decoupled, even in the simple venerable linear model.  

\section{Performance comparisons}
\subsection{Simulation study}
This section considers the improvement the partial factor model can bring over standard Bayesian alternatives:   the conjugate linear model with an independent ``ridge prior" (with unknown ridge parameter) and the Bayesian factor regression model.  We observe via simulation studies that the partial factor model protects against the case where the response loads on a comparatively weak factor.  The partial factor model is most frequently the best performing model (modally optimal), and it is also the best model on average (mean optimal) in unfavorable low signal-to-noise regimes and nearly so in the high signal-to-noise case. In summary, the partial factor model predicts nearly as well as the conjugate linear model and factor models when those models perform well, but it does much better than those models in cases where they do poorly.  This profile is consistent with results of the multiple-shrinkage principal component regression model of \cite{georgeMSPC}, which has a similar motivation -- seeking to mimic principal component regression but to protect against the least-eigenvalue scenario -- but is not derived from a joint sampling model.

For this simulation study, let $p=80$ and $n=50$.   Of the fifty observations, $35$ observations are labeled with a corresponding $Y$ value.  Across 150 data sets, the remaining 15 unlabeled values were predicted using the posterior mean imputed value.  

The data was generated according to the following recipe.
\begin{enumerate}
\item Draw $k \sim \mbox{Uniform}(\{ 1, \cdots, n-1\})$. 
\item Generate a matrix $A$ of size $p \times k$ with independent standard Normal random variables.
\item Generate a $k \times k$ diagonal matrix $D$ with elements drawn from a half-Cauchy distribution.
\item Set the true loadings matrix $B \equiv AD/|AD|$ where the norm is the Frobenius norm.
\item The elements of $\Psi$ are drawn independently as folded-t random variables with 5 degrees of freedom and scale parameter 0.1.
\item Lastly, $\theta$ was drawn by first drawing a folded-t scale parameter and then drawing a mean zero random variable with corresponding scale.
\end{enumerate}

We consider two scenarios. In the first, the elements of $\theta$ and $D$ are ordered so that the highest absolute value of $D$ corresponds to the highest absolute value of $\theta$, the second highest corresponds to the second highest, etc.  This is a favorable case for the assumptions of ridge regression and factor models in that the response depends most on the directions of highest variability in predictor space. For the second case the elements of $\theta$ and $D$ are arranged in reverse, the smallest absolute value of $\theta$ is associated with the largest absolute value of $D$.  In this case the highly informative directions in predictor space are least informative of the response in terms of variation explained.

\begin{table}[h!]
\begin{center}\caption{PFR: Partial factor regression.  NIG:  conjugate prior linear regression.  BFR:  Bayesian factor regression.  Both the factor model and the partial factor model selected $k$ a priori by looking at the singular values of design matrix, so that the top $k$ singular vectors account for 90\% of the observed variance.}
\begin{tabular}{rccc}
{\it Case One.}\\
Method
&  \% best&  mean relative error & scaled MSE\\
 \hline
 PFR&36&.37&1.06\\

NIG&19&.48&1\\

 BFR&29&7.27&1.89\\
 \\
{\it Case Two.}\\
Method\\
\hline
 PFR&43&.27&1\\

 NIG&17&.45&1.04\\

 BFR&30&3.87&1.32\\
 \end{tabular}
\end{center}
\end{table}

To compare the average behavior of these methods on a wide range of data we may look at the paired hold out error on each of the sets.  We record the frequency that each method was the best performing method, the average relative error (the average of the ratio of the squared error of the method to
the minimum squared error over the three methods), and also the overall mean square error.  The first measure records how often we should expect a method to be the best method to use on a randomly selected data set, so that higher numbers are better.  The second column reflects how far off, on average, a given method performs relative to the best method for a given data set; smaller numbers are better.  The final column gives the average error relative to the best overall method;  numbers nearer to one are better.

We observe that in the favorable setting the pure factor model is quite often the best model of the four, as shown in the first column.  However, we notice also that when it is not the best, it performs, on average, much worse than the best method, as shown in the second column.  This is the impact of the bias.  Next, we note that while ridge regression moderately outperforms the partial factor model in terms of overall mean squared error, we see that on average partial factor regression is closer to the best performing model.  Relatedly, it is the partial factor model that is most often the best model.

In the unfavorable setting, results unambiguously favor the partial factor model.  In this setting, as expected, the partial factor model outperforms ridge regression by all three measures.  Again, the pure factor model is crippled by its too-strong bias.   

\subsection{Real data examples}
In this section, we extend our comparisons to additional methods and to the case of real data.  We compare partial factor regression to four other methods:  principal component regression (PCR), partial least squares (PLS), least-angle regression \citep{lars} (LARS), and ridge regression (RR).  We observe the same pattern of robust prediction performance as in the simulation study.  Partial factor regression shows itself to be the best or nearly the best among the methods considered in terms of out-of-sample mean squared prediction error.

Five real data sets in the $p > n$ regime are analyzed; the data are available from the R packages {\tt pls} \citep{pls}, {\tt chemometrics} \citep{chemometrics}, and {\tt mixOmics} \citep{mixomics}.  \begin{quote}
{\bf nutrimouse}: the hepatic fatty-acid concentrations of 40 mice are regressed 
upon the expression of 120 liver cell genes. \\

{\bf cereal}:   the starch content of 15 cereals is regressed upon 145 different wavelengths of NIR spectra. \\

{\bf yarn}:  the yarn density of 28 polyethylene terephthalate (PET) yarns is regressed upon 268 wavelenths of NIR spectra. \\

 {\bf gasoline}:  the octane numbers of 60 gasoline samples are regressed upon 401 wavelengths of NIR spectra. \\

{\bf multidrug}:  an ATP-binding cassette transporter (ABC3A) is regressed upon the the activity of 853 drugs for 60 different human cell lines.
\end{quote} 

To test the methods, each of the data sets is split into training and test samples, with 75\% of the observations used for training. Each model is then fit using the training data, with tuning parameters chosen by ten-fold cross validation on only the training data. Out-of-sample predictive performance on the holdout data is measured by sum of squared prediction error (SSE).

\begin{table}[h!]
\begin{center}\caption{PFR: partial factor regression.  RR:  ridge regression.  PLS:  partial least squares.  LARS:  least angle regression.  PCR:  principal component regression. Percentages shown are amount worse than the best method, reported in bold type.}
\hspace{1.25in} Average out-of-sample error\\
\begin{tabular}{rcclllll}\label{dataset}
Data set & $n$ & $p$ & PFR & RR & PLS & LARS & PCR\\
 \hline
nutrimouse& 40&100&435.0 (4\%) &{\bf 418.72}&448.3 (7\%)&502.3 (20\%) &454.2 (8\%) \\
cereal&15&120&{\bf 44.4} & 49.5 (11\%) & 51.2 (15\%) & 69.0 (55\%) & 54.3 (22\%)\\
yarn&28&145&{\bf 0.16} & 0.47 (194\%) & 0.47 (194\%) & 0.39 (144\%) & 0.58 (263\%)\\
gasoline&60&269&{\bf 0.68}&0.79 (16\%)&0.86 (27\%)&1.04 (52\%)&0.80 (18\%)\\
multidrug&60&401&167.6 (6\%) & {\bf 158.8} & 159.9 (1\%) & 198.1 (25\%) & 167.8 (6\%)\\
 \end{tabular}
\end{center}
\end{table}

As shown in table \ref{dataset}, the partial factor model outperforms all models on three of the five data sets.  In the other two data sets,  the {\tt nutrimouse} and {\tt multidrug} examples, the factor structure was weak, requiring $k \approx n$ to account for the variation in the predictor space.  In these cases, the extra variance of learning two tuning parameters does not pay dividends and ridge regression narrowly comes out on top.  Even so, partial factor regression is never much worse than the best.  In cases where the predictor space can be described well by low dimensional (linear) structure, partial factor regression out-performs methods such as principal component regression, which require that this same structure account for all of the variability in the response.  

Note that these data were selected because they are publicly available and fall within the $p > n$ regime that is most germane to our comparisons. 

\section{Variable selection and subspace dimension estimation}
\subsection{Sparsity priors for variable selection}
In this and the next section, it is convenient to work with a reparametrized form of the partial factor model, defining \begin{equation} \Lambda = (V - \theta B^t)\Psi^{-\frac{1}{2}} \end{equation} and using the equivalent independent prior \begin{equation} \Lambda \sim \N(0, w^{-1}\I_p). \end{equation} Note that $\Lambda = 0$ represents a pure factor model, and that this prior is independent of the other parameters.  The revised expression for our (latent) regression becomes
\begin{equation}
Y  = \theta f+\Lambda\Psi^{-\frac{1}{2}}(X- B f) + \epsilon, \hspace{0.2in} \epsilon \sim \N(0, \sigma^2).
\end{equation}

If $\lambda_j = 0$,  predictor $X_j$ appears in the regression of $Y$ only via its dependence on the latent factors.  Further, if we assume that $\theta$ is not identically zero so that $Y$ has some relation to the latent factors, then we see that  if $b_j = 0$ (so that dimension $j$ does not load on any of the factors) and $\lambda_j=0$, then $\beta_j = 0$ necessarily.  That is, if $X_j$ is not related to any of the latent factors governing the predictor covariance and additionally is not idiosyncratically correlated with $Y$ via $\lambda_j$, then $X_j$ does not feature in our regression.  The reverse need not hold; the net effect of $X_j$ on $Y$ can appear insignificant if $X_j$  has a direct effect on the response, but is positively correlated with variables having the opposite effect. 

Partial factor regression helps distinguish between these two scenarios, because the framework permits sparsity to be incorporated in each of three separate locations, with the following easy interpretations.  \begin{enumerate}
\item Does variable $X_j$ load on latent factor $f_g$?  ($b_{jg} = 0$ versus $b_{jg} \neq 0$ )
\item Does $Y$ depend on the residual of element $X_j$; is $X_j$ important for predicting $Y$ above and beyond the impact of the latent factors?  ($\lambda_j =  0$ versus $\lambda_j \neq 0$)
\item Does $Y$ depend on latent factor $f_g$?  ($\theta_g = 0$ versus $\theta_g \neq 0$)
\end{enumerate}

This decomposition avoids the unsatisfactory choice of having to decide which of two variables should be in a model if they are very highly correlated with one another and associated with the response.  Rather it allows one to consider the common effect of two such variables in the form of a latent factor, and then to consider separately if both or neither should enter into the model residually via the parameter $\Lambda$.  Earlier work has keyed on to the idea that covariance regularization is useful for variable selection problems \citep{jeng}; here these intuitive decompositions follow directly from the generative structure of the partial factor model.

Such a variable selection framework may be implemented with the usual variable selection point-mass priors on $\theta$, $\Lambda$ and $B$.  Previous work incorporated such priors for the elements of $B$ \citep{carlosJASA}.  Alternatively, shrinkage priors may and thresholding may be used to achieve a similar effect.

\subsection{Subspace dimension estimation}
In the case of multivariate Normal random variables, a factor decomposition of the covariance matrix, in combination with point mass priors as described above, admits a ready characterization of the dimension reduction subspace \citep{cookfish, cookforzani, sayanmanifold} with respect to the response $Y$.  A dimension reduction subspace is the span of a projection of the predictors which is sufficient to characterize the conditional distribution of the response.

In the factor model setting, we can calculate the dimension of this subspace as follows \citep{kaiAISTAT}. Let $\theta_Y$ denote the nonzero elements of $\theta$ in the partial factor parameterization.  Denote by $B_Y$ the corresponding columns of $B$ and likewise let $B_X$ denote the remaining columns.  Then, if $\Lambda = 0$, the conditional distribution of $Y$ given $X$ can be characterized purely in terms of 
\begin{equation}
\begin{split}
E(Y \mid X) &=\theta B^t(BB^t+\Psi)^{-1}X\\
&=\theta_Y B^t_T(B_YB^t_Y+B_XB^t_x+\Psi)^{-1}X\\
&\equiv \theta_Y B^t_Y(B_YB^t_Y+\Delta)^{-1}X\\
&=\theta_Y [ \mbox{I} - B^t_Y\Delta^{-1}B_Y(\mbox{I} + B^t_Y\Delta^{-1}B_Y)^{-1)}] B^t_Y\Delta^{-1}X,
\end{split}
\end{equation}
where $\Delta \equiv B_X B_X^t+ \Psi$, showing that $X$ enters this distribution only via $B^t_X\Delta^{-1}X$.  Thus, the rank of $B_Y$ is the dimension of the reduced subspace, as long as we have a pure factor model.  We have already seen, however, that while a covariance matrix may be relatively well approximated by a small number of factors, these factors alone may not span the dimension reduction subspace, so that $\theta$ is estimated to be approximately zero and $\sigma^2$ is biased upward.   

Accordingly, we estimate $\mbox{Pr}(\Lambda = 0 \mid X, Y)$, the posterior probability that the sufficient subspace is less than $k = \mbox{rank}(B)$.  Further, by monitoring the number of nonzero elements of $\theta$ in our sampling chain, we can estimate the sufficient dimension, conditional on it being less than $k$.  This approach may be thought of as partitioning our prior hypotheses as
\begin{equation}
\begin{split}
\mathcal{H}_j &: \mbox{rank}(B_Y) = j, \hspace{.1in} j \in \lbrace 1, \dots, k \rbrace \\
\mathcal{H}_0 &: \mbox{rank}(B_Y) > k. 
\end{split}
\end{equation}
The prior probabilities assigned to these hypotheses are induced via priors on $k$ and $\Lambda$; grouping many individual hypotheses into the aggregate $\mathcal{H}_0$ permits easier control of the contribution of the prior, which can be critical to inference when $p \gg n$.

\section{Conclusions}
In the $p \gg n$ setting, inference and prediction may sometimes be improved by making structural simplifications to the statistical model.  In a Bayesian framework this can be accomplished by positing lower dimensional latent variables which govern the joint distribution between predictors and the response variable.  An inherent downside to this approach is that it requires specifying a high dimensional joint sampling distribution and the associated priors.  Due to the high dimensionality this task is difficult, particularly with respect to appropriately modulating the implied degree of regularization of any given conditional regression.  

The partial factor model addresses this difficulty by reparametrizing the joint sampling model using a compositional representation, allowing the conditional regression to be handled independently of the marginal predictor distribution.  Specifically,  this formulation of the joint distribution realizes borrowing of information via a hierarchical prior rather than through a fixed structure imposed upon the joint distribution. 

Here we have examined the simplified setting of a joint Normal distribution.  However, the idea of utilizing a compositional representation in conjunction with a hierarchical prior can be profitably extended to many joint distributions.  In particular, one may specify the joint distribution directly, building in borrowing of information by design.  For example, the form of the conditional moment for the partial factor model suggests the following nonlinear generalization:
\begin{equation}
E(Y \mid f, X) = g(f) + h(X - E(X \mid f)),
\end{equation}
where perhaps $g$ and $h$ denote smooth functions to be inferred from the data.  Here, the smoothness assumptions for $g$ and $h$ could be different; specifically the prior on $h$ could be conditioned on properties of $g$.  More generally, the partial factor model is a special case of models of the form:
\begin{eqnarray}
f(Y, X \mid \Theta) &=&f(X \mid \theta_X)f(Y \mid X, \theta_X, \theta_Y)\\
\pi(\Theta) &=& \pi(\theta_Y \mid \theta_X)\pi(\theta_X),\label{prior}
\end{eqnarray}
where $\Theta = \lbrace \theta_X, \theta_Y \rbrace$ generically denotes parameters governing the joint distribution.  In the partial factor model $\theta_X = \lbrace B, F, \Psi \rbrace$ and $\theta_Y = \lbrace \sigma^2, V \rbrace$.  The conditional model depends on {\it both} $\theta_X$ and $\theta_Y$, but the presence of $\theta_Y$ in the model leads to a more flexible regression, while the hierarchical prior (\ref{prior}) still borrows information from the predictor variables via $\theta_X$.

Such models alleviate the burden of having to get the high dimensional distribution just right in all of its many details.  As such, it represents a robust method for fashioning data-driven prior distributions for regression models.

\bibliographystyle{abbrvnat}
\bibliography{pfr}
\end{document}